\documentclass[final,1p,times]{elsarticle} 
\usepackage{graphicx}
\usepackage{amssymb} 
\usepackage{amsthm} 
\usepackage{lineno}
\usepackage{color}

\journal{Nuclear Physics A} 

\begin{document}

\begin{frontmatter} 

% Your Title - please insert
\title{Mixed harmonic charge dependent azimuthal correlations in Pb-Pb
  collisions at $\sqrt{s_{NN}}=2.76$ TeV measured with the ALICE experiment at the LHC}

%% Single author (and collaboration) - please insert
\author{Yasuto Hori (for the ALICE\fnref{col1} Collaboration)}
\fntext[col1] {A list of members of the ALICE Collaboration and acknowledgements can be found at the end of this issue.}
\address{ Center for Nuclear Study, Graduate School of Science, University
  of Tokyo, Japan}

%% Multiple authors
%\author[auth2]{Marcus Junius Brutus}
%\address[auth1]{Somewhere, Rome}
%\address[auth2]{Somewhere else, Rome}

\begin{abstract} 
Mixed harmonic charge dependent azimuthal correlations at mid-rapidity in
Pb-Pb collisions at $\sqrt{s_{NN}}=2.76$ TeV were measured with the ALICE detector
at the LHC. A clear charge dependence for a series of correlations is observed both via the
multi-particle cumulant and the event plane methods. 
Implications from these measurements for the possible
effects of the local parity violation in QCD and for models
which incorporate the azimuthal anisotropic flow and the local charge conservation on
the kinetic freeze-out surface are discussed. 
\end{abstract} 

\end{frontmatter} % do not change

%% linenumbers are useful for reviewing process
%%\linenumbers

\vspace{-2mm}
\section{Introduction}
%The charge dependence of the azimuthal correlations between produced
%hadrons is an important probe of the QGP matter created in 
%relativistic heavy-ion collisions. 
%It has been pointed out that the local parity violations in QCD are
% not forbidden and may metastably
% happen as a result of topologically non-trivial gluon configurations during heavy ion collisions. 
%These possible parity odd bubbles have the difference in the number of quarks with positive
%and negative chiralities. Therefore, combined with a large magnetic field perpendicular to the reaction plane generated during
%non-central heavy ion collisions, it is expected that the existence of
%these bubbles gives rise to the charge separation with respect to the reaction
%plane. This possible phenomenon is called the Chiral Magnetic Effect
%(CME) ~\cite{a02}.

The charge dependence of the azimuthal correlations
between produced hadrons is an important probe
of the QGP matter created in relativistic heavy-ion collisions.
It is in particular sensitive to the interplay between the
local charge conservation (LCC) induced correlations and azimuthally asymmetric radial expansion
of the collision system \cite{a05}.
This further helps to discriminate effects from charges produced earlier
in the collision by the gluon string fragmentation and late in the collision
by the hadronization of the expanding QGP matter.\\
Recently, it was argued that the charge dependent azimuthal
correlations can be also sensitive to the possible effect of the local parity violation in QCD \cite{a02}.
Parity violation in QCD may happen as a result of the interaction between
produced quarks and topologically non-trivial gluonic field configurations.
In the presence of the strong magnetic field generated in a heavy-ion collision,
the local parity violation may result in a separation of charges along the magnetic field
which points perpendicular to the reaction plane.
This phenomenon is called the Chiral Magnetic Effect (CME).\\
An important observable as a sensitive probe of the
CME is the two particle correlation with respect to the reaction plane
$\langle
{\rm cos}(\varphi_{\alpha}+\varphi_{\beta}-2\Psi_{RP})\rangle$ \cite{a03},
where the bracket denotes the average over all particles in all events
and the indices $\alpha$ and $\beta$ refer to the charge of the
particles.
$\varphi_{\alpha, \beta}$ is the azimuthal angle of the charged particles
and $\Psi_{RP}$ is the reaction plane angle. 
Measurements by the STAR Collaboration revealed
non-zero correlation $\langle
{\rm cos}(\varphi_{\alpha}+\varphi_{\beta}-2\Psi_{RP})\rangle$, which is consistent
with qualitative expectations from the CME \cite{a04}.
At the same time, a study ~\cite{a05} showed
that a significant part of the observed charge dependence can be
described by the Blast Wave model incorporating effects of the LCC.
On the other hand, the charge independent part of this correlation may have 
contributions from the dipole flow fluctuations and effects
of momentum conservation \cite{a09, a10}.
At the moment, none of the models can reproduce simultaneously
the charge dependent and independent parts of the observed correlations.
Recently the ALICE Collaboration released a paper \cite{a07}
where the correlation $\langle
{\rm cos}(\varphi_{\alpha}+\varphi_{\beta}-2\Psi_{RP})\rangle$ were
measured at LHC energies.
In these proceedings, we extent the ALICE measurement
to the additional mixed harmonic charge dependent correlations with
respect to the collision symmetry plane, which may help to disentangle
the CME and LCC induced correlations [9].

\vspace{-2mm}
\section{Analysis details}
A sample of about 13 M minimum bias Pb-Pb collisions at
$\sqrt{s_{NN}} = 2.76$ TeV
collected by the ALICE detector during the 2010 LHC run was analyzed.
Description of the ALICE detector and details about collision triggers
and centrality determination
can be found in \cite{a07, ALICE-PPR}.
%Analyzed data sample is $\sim13$~M minimum bias $\sqrt{s_{NN}}=2.76$~TeV
%Pb-Pb events. 
%For the $\alpha$ and $\beta$ particles, tracks reconstructed by the Time
%Projection Chamber (TPC) at $|\eta|<0.8$ and $p_{T}>0.2$~GeV/$c$ are used.
A Time Projection Chamber (TPC) is used to reconstruct charged particles
in the kinematic range $|\eta| < 0.8$ and $p_{T} > 0.2$ GeV/$c$.
Correlations with respect to the symmetry plane
were measured
using the event plane and multi-particle cumulant methods.
In the event plane method, the symmetry planes were estimated
from azimuthal distributions of hits in two forward scintillator
counters (VZERO)
which cover the pseudo-rapidity range $-3.7 <  \eta < -1.7$ and $2.8 <
\eta  < 5.1$,
and two Forward Multiplicity Detectors (FMD) located at $1.7 < \eta <
5.1$ and $-3.4 < \eta < -1.7$.
In the multi-particle cumulant method, the correlations with respect
to the symmetry plane are evaluated from the azimuthal angle correlations of charged
particles reconstructed by the TPC.
Although the dominant systematic errors come from the
  event plane determinations,
  we observed good agreement between results from two different methods.
%A small differences between results were used to evaluate systematic
%uncertainties \textcolor{red}{from the event plane determinations}.

%A small differences between results were used to evaluate systematic
%uncertainties.

%To calculate the correlation $\langle {\rm cos} [n\varphi_{\alpha} +
%  m\varphi_{\beta}-(n+m)\Psi_{k}] \rangle$, two independent methods - event
%plane method and multi-particle cumulant method - are conducted and
%the agreement between these methods was confirmed. In the
%event plane method, the symmetry plane $\Psi_{k}$ is estimated through the
%event plane calculated by hits at the forward detectors: VZERO detector at $-3.7<\eta<-1.7$ and $2.8<\eta<5.1$
%consists of plastic scintillation counters and Forward Multiplicity
%Detector (FMD) at $1.7<|\eta|<5.1$ consists of silicon strip
%detectors. In the multi-particle cumulant method, the symmetry
%plane is indirectly estimated by azimuthal angles of another particles.
%\begin{eqnarray}
%\langle {\rm cos}[n\varphi_{\alpha}+m\varphi_{\beta}-(n+m)\Psi_{k}] \rangle \sim
%\frac{\langle {\rm cos}[n\varphi_{\alpha}+m\varphi_{\beta} \dots -
%  k\varphi_{\gamma_{(n+m)/k}}] \rangle_{c} }{ v_{k, \gamma_{1}} \times \dots
%\times v_{k, \gamma_{(n+m)/k}}} \nonumber
%\end{eqnarray}
%where the bracket in the numerator denotes the multi-particle cumulant. In the
%denominator the average of $v_{k}$\{$2$\} and $v_{k}$\{$4$\} is used
%to have $v_{k}$.
 \vspace{-2mm}
\section{Results}
The centrality dependence of the correlations $\langle
{\rm cos}(\varphi_{\alpha}-\varphi_{\beta})\rangle$ and $\langle
{\rm cos}(\varphi_{\alpha}+\varphi_{\beta}-2\Psi_{RP})\rangle$ for the same
and opposite charge combinations measured for Pb-Pb collisions
at $\sqrt{s_{NN}} = 2.76$ TeV was reported by ALICE in [6].
For the correlation $\langle
{\rm cos}(\varphi_{\alpha}+\varphi_{\beta}-2\Psi_{RP})\rangle$, ALICE observed that
the same charge correlation is non-zero and negative, while the opposite charge
correlation has a significantly smaller magnitude and is positive for peripheral collisions.
ALICE also showed that there is little collision energy dependence when comparing 
results to that at the top RHIC energy.
Even though some of the features of the observed correlations are in qualitative
agreement with the expectation from the CME,
the origin of these correlations is still not clear since they are sensitive to
many other parity-conserving physics mechanisms.
To study the physics backgrounds for the CME search, ALICE has measured the two
particle correlation $\langle
{\rm cos}(\varphi_{\alpha}-\varphi_{\beta})\rangle$, which also shows strong charge dependence but its
correlation strength is significantly different from what was measured
by the STAR Collaboration
at lower collision energy.
As shown in Fig.1 and 2, measurements were extended to a set of charge
dependent correlations $\langle {\rm cos} [\varphi_{\alpha} -
  (m+1)\varphi_{\beta}+m\Psi_{k}] \rangle$ where $m,k$ are integers and
$\Psi_{k}$ is the $k$-th order collision symmetry plane angle.
%Here an approximation $\Psi_{RP}\sim\Psi_{2}$ is assumed.
They help to better constrain the possible physical contributions
to the previously measured correlation $\langle {\rm
  cos}(\varphi_{\alpha}+\varphi_{\beta} - 2\Psi_{RP})\rangle$.
% since the LCC induced correlations affect not only the
%correlation $\Delta \langle {\rm
%  cos}(\varphi_{\alpha}+\varphi_{\beta} - 2\Psi_{RP})\rangle$ but also
%on the other correlations $\Delta \langle {\rm cos} [\varphi_{\alpha} -
%  (m+1)\varphi_{\beta}+m\Psi_{k}] \rangle$ as the combined effects with the
%azimuthal anisotropic flow $v_{m}$\{$\Psi_{k}$\} where $\Delta$
%donotes the difference between the same and opposite charge correlations}.
Note that the charge independent baseline correlation $\langle {\rm cos}(\varphi_{\alpha}-3\varphi_{\beta} +
2\Psi_{2})\rangle$  is related
to the widely discussed effects of flow fluctuations and inter-correlation
between the collision symmetry planes \cite{a09, Ante-QM2012-proceedings}.\\
%We extend the set of measured correlations as shown in Figure 1 and 2.
%\begin{figure}[hh]
%  \centering  \includegraphics[width= 6.5cm]{2012-Aug-08-C1c3_2.eps}%, height=4.5cm]{pub2p.eps}
%  \\
\begin{figure}[hh]
  \begin{tabular}{cc}
    \begin{minipage}{0.5\hsize}
      \centering
      \includegraphics[width= 6.5cm]{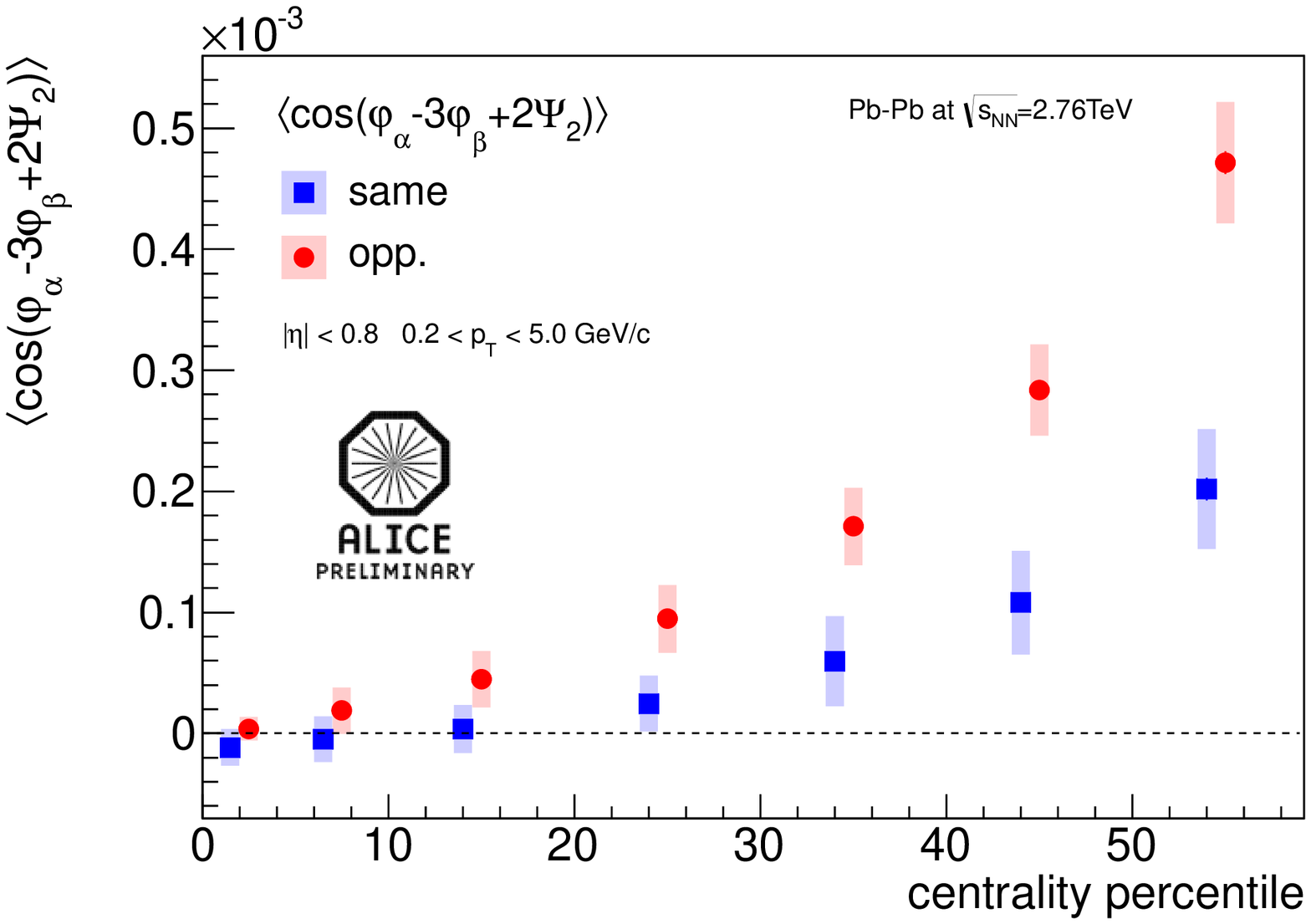}%{2012-Aug-08-C1c3_2.eps}%, height=4.5cm]{figure2.eps}
    \end{minipage}
    \begin{minipage}{0.5\hsize}
      \centering
      \includegraphics[width= 6.5cm]{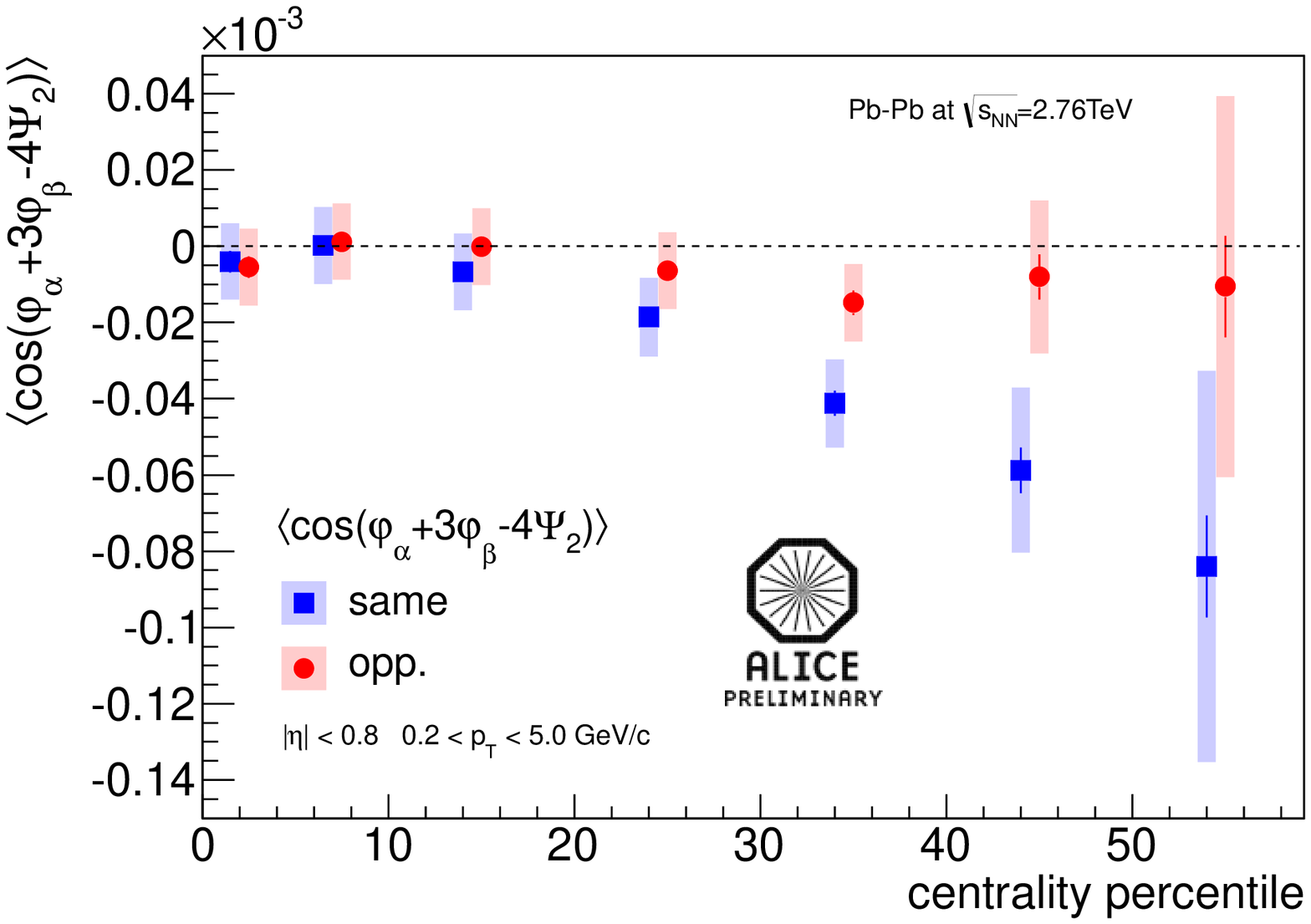}%{2012-Aug-08-C13_2_0.eps}%, height=4.5cm]{figure2.eps}
    \end{minipage}
  \end{tabular}
  \caption{Centrality dependence of the charge dependent two particle
  azimuthal correlations
  with respect to the 2nd order symmetry plane $\Psi_{2}$: (left)
  $\langle {\rm cos}(\varphi_{\alpha}-3\varphi_{\beta}+2\Psi_{2}) \rangle$, (right) $\langle {\rm cos}(\varphi_{\alpha}+3\varphi_{\beta}-4\Psi_{2}) \rangle$.}
%\end{figure}
%\vspace{-3mm}
%These additional measurements allows to better constrain the possible
%physical contributions
%to the previously measured $\langle {\rm cos}(\varphi_{\alpha}+\varphi_{\beta} -
%2\Psi_{2})\rangle$.
%\begin{figure}[hh]
  \begin{tabular}{cc}
    \begin{minipage}{0.5\hsize}
      \centering
      \includegraphics[width= 6.5cm]{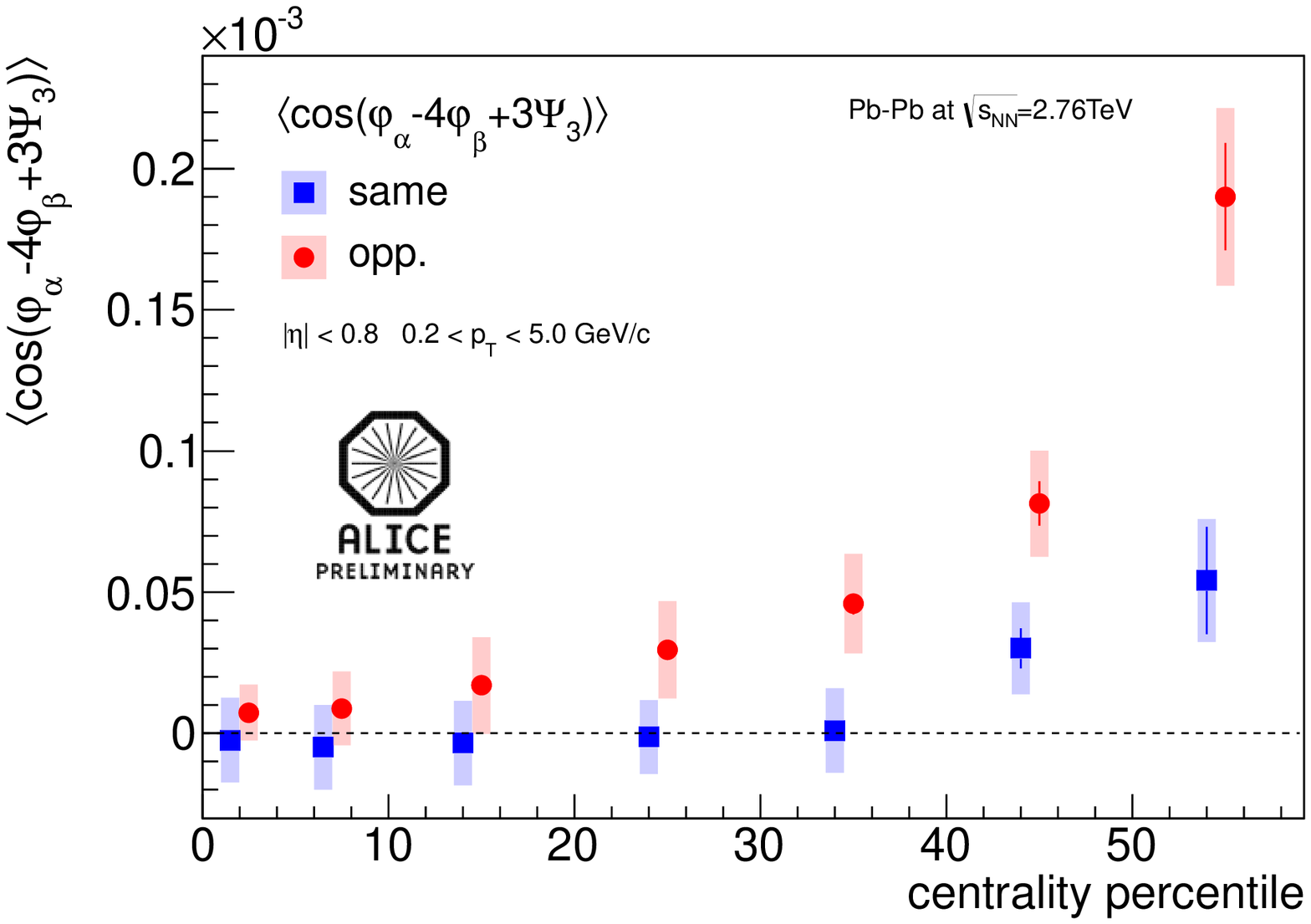}%{2012-Aug-08-C1c4_3.eps}%, height=4.5cm]{pub2p.eps}
    \end{minipage}
    \begin{minipage}{0.5\hsize}
      \centering
      \includegraphics[width= 6.5cm]{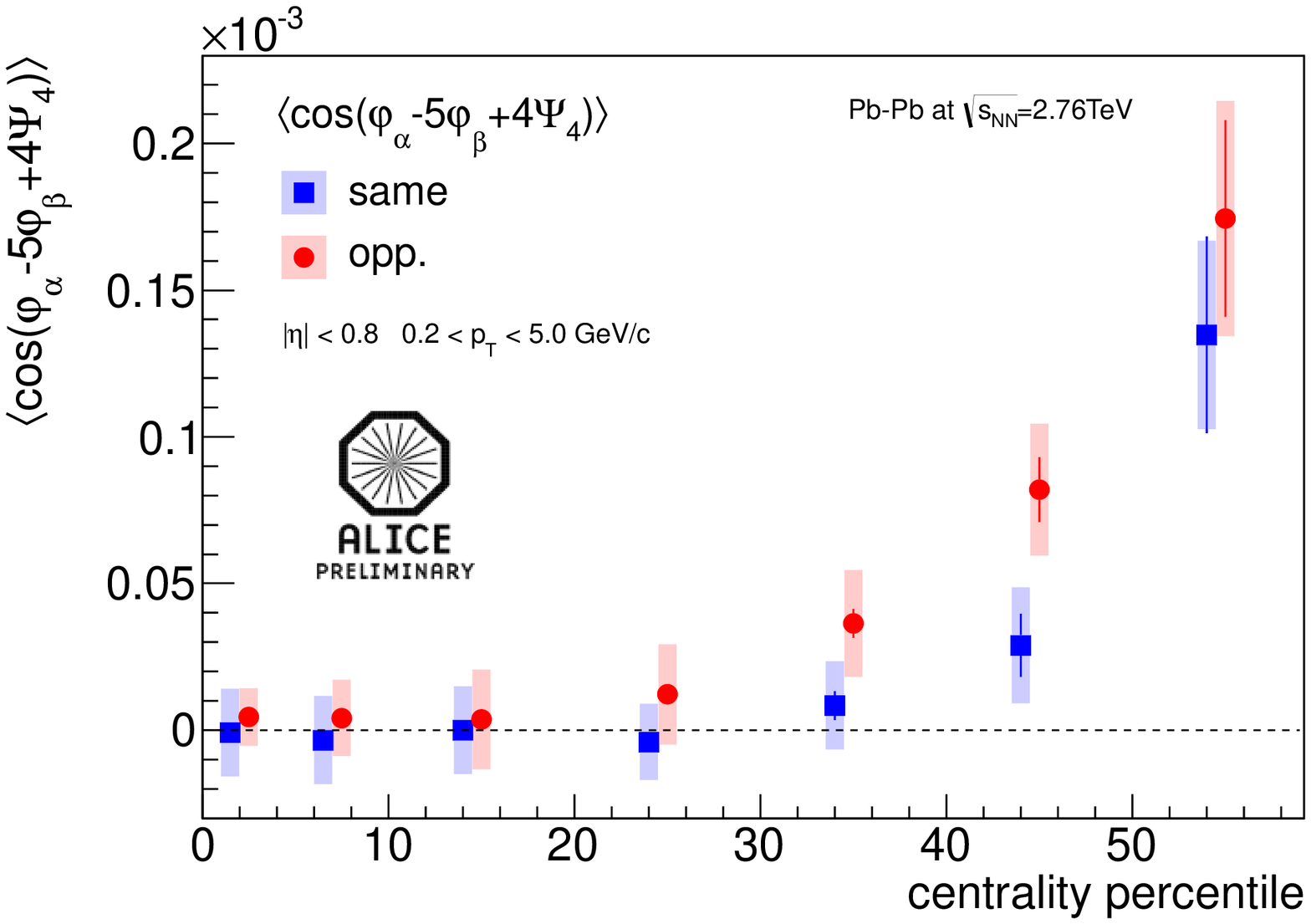}%{2012-Aug-08-C1c5_4.eps}%, height=4.5cm]{figure2.eps}
    \end{minipage}
  \end{tabular}
  \caption{Centrality dependence of the charge dependent two particle
  azimuthal correlations
  with respect to the 3rd and 4th order symmetry plane $\Psi_{3,4}$: (left)
  $\langle {\rm cos}(\varphi_{\alpha}-4\varphi_{\beta}+3\Psi_{3}) \rangle$, (right) $\langle {\rm cos}(\varphi_{\alpha}-5\varphi_{\beta}+4\Psi_{4}) \rangle$.}
\end{figure}
%\vspace{-3mm}
%\begin{figure}[hh]
%  \begin{tabular}{cc}
%    \begin{minipage}{0.5\hsize}
%      \centering
%      \includegraphics[width=6.5cm]{pub2p.eps}%, height=4.5cm]{pub2p.eps}
%    \end{minipage}
%    \begin{minipage}{0.5\hsize}
%      \centering
%      \includegraphics[width=6.5cm]{figure2.eps}%, height=4.5cm]{figure2.eps}
%    \end{minipage}
%  \end{tabular}
%  \caption{Left: the centrality dependence of the correlation $\langle
%{\rm cos} (\varphi_{\alpha} - \varphi_{\beta}) \rangle$ measured by the
%ALICE and STAR collaboration, Right:  the centrality dependence of the correlation $\langle
%{\rm cos} (\varphi_{\alpha} + \varphi_{\beta} - 2\Psi_{RP}) \rangle$ measured by the
%ALICE and STAR collaboration.}
%\end{figure}
%\vspace{-3mm}
%Figure 3 shows the charge dependent part of the
%correlations $\langle
%{\rm cos} [n(\varphi_{\alpha} - \varphi_{\beta})] \rangle$ (left) and $\langle {\rm
%  cos} [\varphi_{\alpha}-(m+1)\varphi_{\beta}+m\Psi_{2}] \rangle$ (right)
%in comparison with the Blast Wave model incorporating effects of the LCC \cite{a06}.
The model in Fig.3 with parameters tuned on the measured
hadron spectra and the anisotropic flow at the LHC
 reproduces the features of the charge dependence
of the correlations $\langle
{\rm cos} (\varphi_{\alpha} - \varphi_{\beta}) \rangle$ and $\langle {\rm
  cos} [\varphi_{\alpha}-(m+1)\varphi_{\beta}+m\Psi_{2}] \rangle$, but fails to describe
the higher moments $\langle
{\rm cos} [n(\varphi_{\alpha} - \varphi_{\beta})] \rangle$ for $n>$1 \cite{a06}.
This indicates that a significant part of the correlations observed
for $\langle
{\rm cos}(\varphi_{\alpha}+\varphi_{\beta}-2\Psi_{RP})\rangle$
may originate from the LCC induced correlation. However, more
studies are needed
to quantify its actual contribution.
ALICE also measured
a charge dependence of the double-harmonic correlation $\langle {\rm cos}(2\varphi_{\alpha}+2\varphi_{\beta}- 4\Psi_{4})\rangle$
which also may help in disentangling effects from the LCC and
the CME  \cite{S.Voloshin-QM2012-proceedings}.\\
Similarly to the differential correlations for
$\langle {\rm cos} (\varphi_{\alpha}+\varphi_{\beta}-2\Psi_{RP}) \rangle$
reported in \cite{a07}, ALICE observes that the
other correlations $\langle {\rm cos} (\varphi_{\alpha}-3\varphi_{\beta}+2\Psi_{RP}) \rangle$ 
are localized within about one unit of rapidity (or may even change sign as a
function of $\Delta\eta$)
and extend up to the higher $p_{T}$ of the pair as shown in Fig.4.
\vspace{-2mm}
\section{Summary}
%We observed a clear charge dependence of the correlation $\langle {\rm cos}
%(\varphi_{\alpha}+\varphi_{\beta}-2\Psi_{RP}) \rangle$ which is consistent
%with the CME expectations. However, a series of the mixed harmonic
%azimuthal correlations 
%including the correlation $\langle {\rm cos}
%(\varphi_{\alpha}+\varphi_{\beta}-2\Psi_{RP}) \rangle$ also shows a significant charge
%dependence which can be systematically interpreted as a consequence of
%the anisotropic flow and the strong spatial correlations between the hadron and the anti-hadron
%on the kinematic freeze-out surface. 
Charge dependent azimuthal correlations 
in Pb-Pb collisions at $\sqrt{s_{NN}}=2.76$ TeV were measured by the ALICE Collaboration.
A significant non-zero correlation 
$\langle {\rm cos}(\varphi_{\alpha}+\varphi_{\beta}-2\Psi_{RP})
\rangle$ was observed, which
was originally proposed as an observable sensitive to the CME and thus to effects from the local parity violation in QCD.
The experimental analysis was extended to the higher moments of the
two particle azimuthal correlations
$\langle {\rm cos} [n(\varphi_{\alpha}-\varphi_{\beta})] \rangle$ for $n=$1-4 and to the mixed harmonic
charge dependent azimuthal correlations with respect to the 2nd, 3rd, and 4th order collision symmetry planes
(e.g. $\langle {\rm cos}(\varphi_{\alpha}-3\varphi_{\beta}+2\Psi_{2}) \rangle$).
These new results provide an important experimental input which is relevant
to the study of various physics mechanism responsible for the charge dependence
of the azimuthal correlations among particles produced in a heavy-ion collision,
such as  the CME, local charge conservation, and flow fluctuations.
\begin{figure}[hh]
  \begin{tabular}{cc}
    \begin{minipage}{0.5\hsize}
      \centering
      \includegraphics[width=6.5cm]{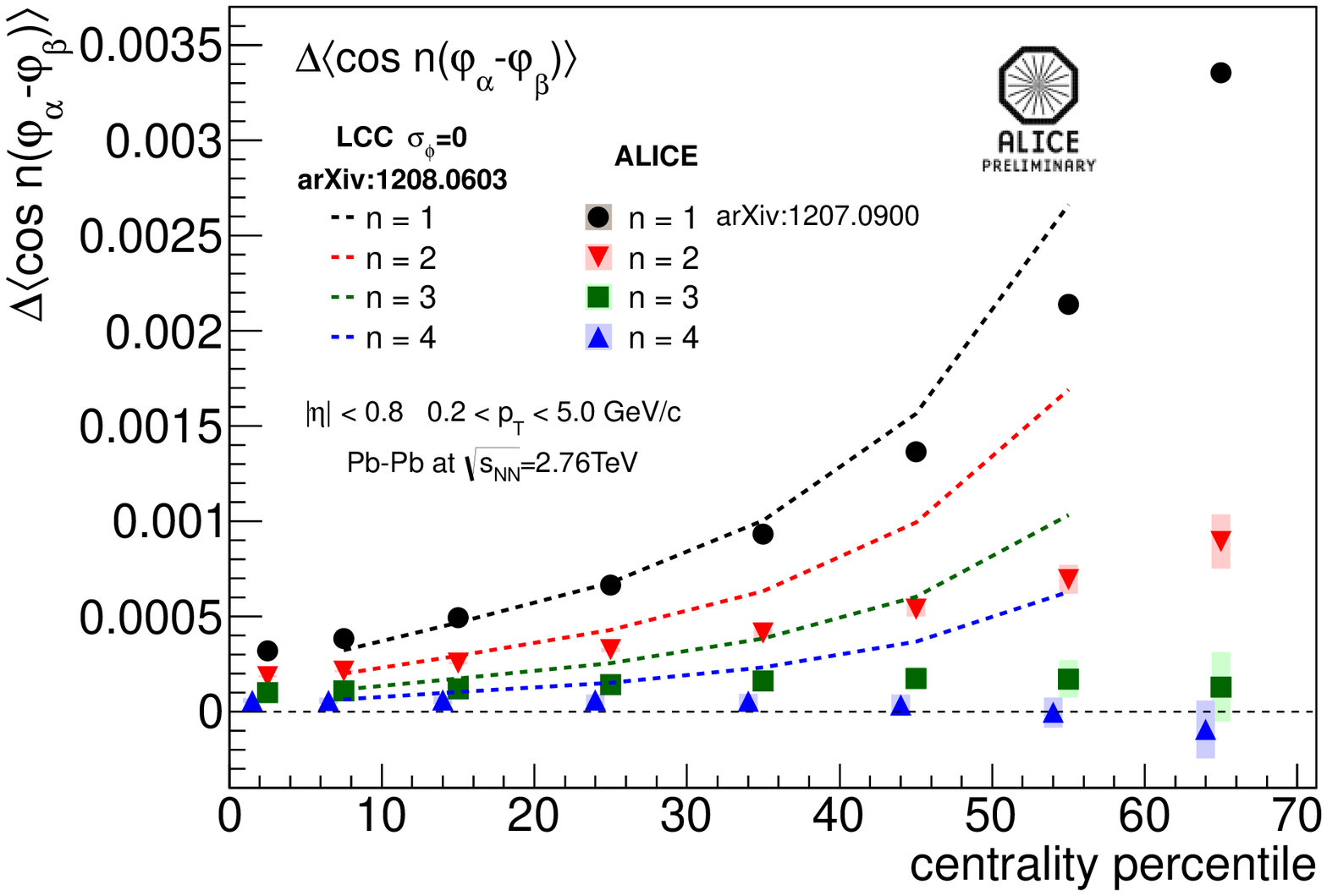}%{2012-Aug-08-dcosn.eps}%, height=4.5cm]{pub2p.eps}
    \end{minipage}
    \begin{minipage}{0.5\hsize}
      \centering
      \includegraphics[width=6.5cm]{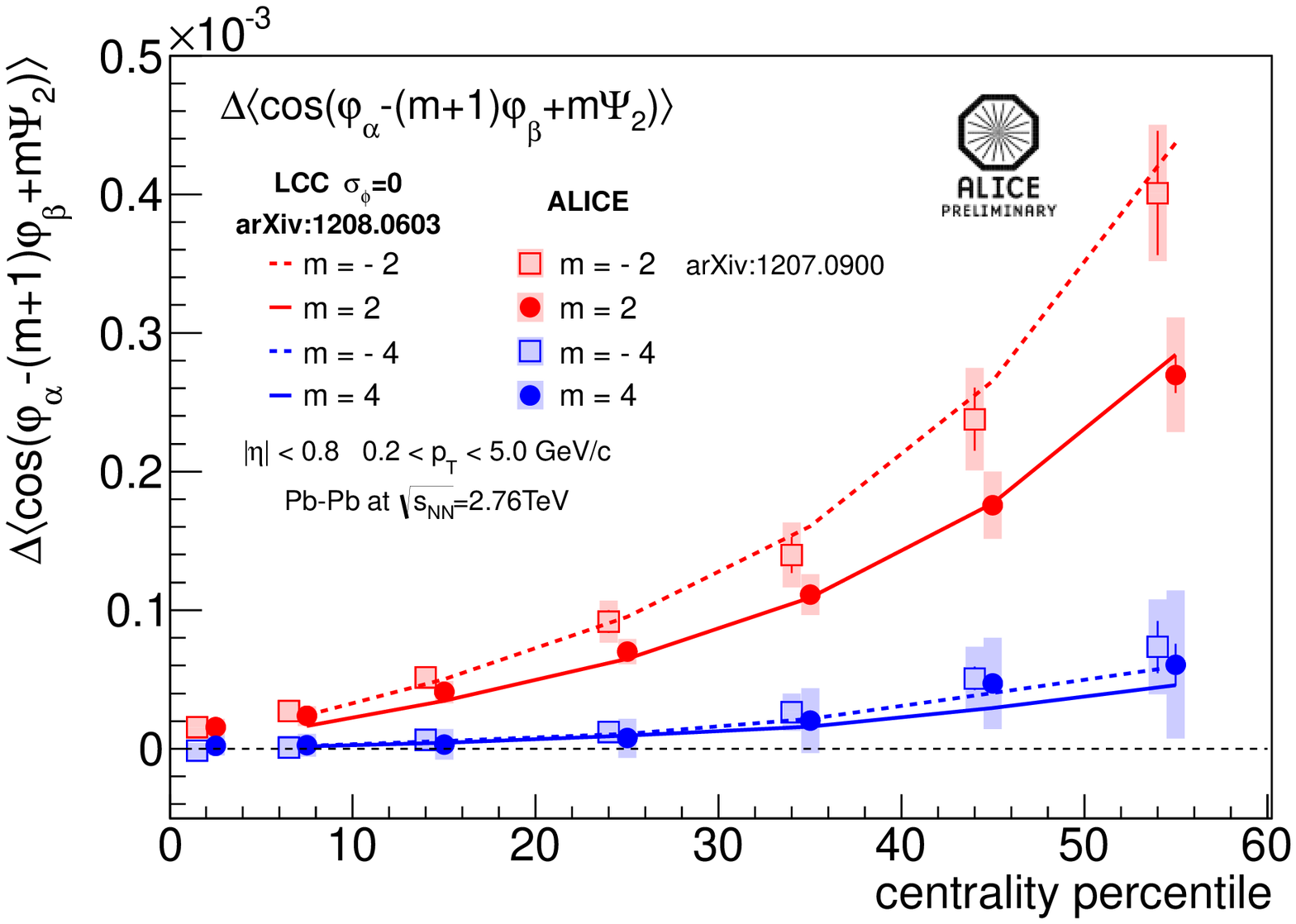}%{2012-Aug-08-DC1mk.eps}%, height=4.5cm]{figure2.eps}
    \end{minipage}
  \end{tabular}
  \caption{Centrality dependence of the correlation $\Delta \langle
{\rm cos} [n(\varphi_{\alpha} - \varphi_{\beta})] \rangle$ (left) and $\Delta \langle {\rm
  cos} [\varphi_{\alpha}-(m+1)\varphi_{\beta}+m\Psi_{2}] \rangle$
(right) in comparison with the Blast Wave model
  incorporating effects of the LCC \cite{a06}.
Here $\Delta$ denotes the difference between the same and opposite
charge correlations.}
%\end{figure}
%\begin{figure}[hh]
%  \centering
%  \includegraphics[width=10cm]{figure3.eps}
  \centering
  \includegraphics[width=10cm]{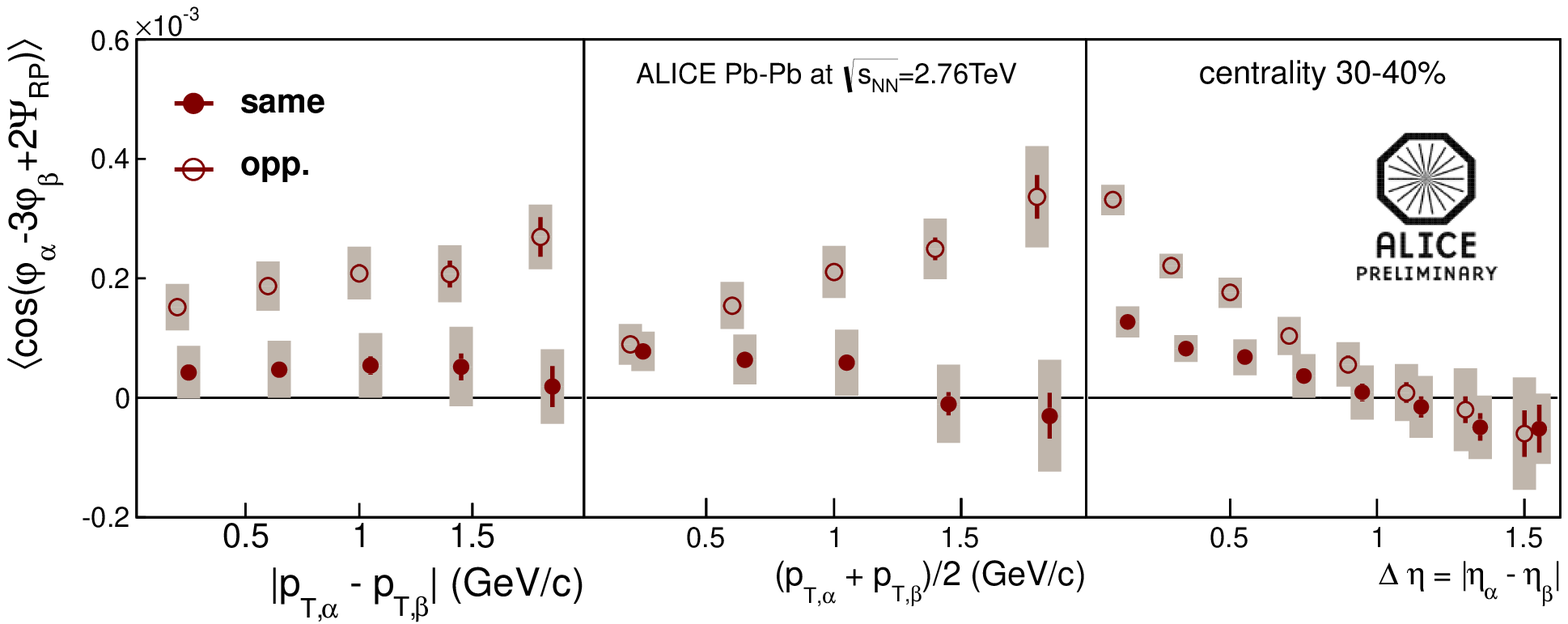}%{2012-Aug-08-diffTY3040.eps}
  \caption{The pair differential correlation 
%$\langle {\rm
%  cos} (\varphi_{\alpha}+\varphi_{\beta}-2\Psi_{RP}) \rangle$ (top) and
$\langle {\rm
  cos} (\varphi_{\alpha}-3\varphi_{\beta}+2\Psi_{RP}) \rangle$ as a function
  of (left) the transverse momentum difference
  $|p_{T,\alpha}-p_{T,\beta}|$,
(center) the average transverse momentum $(p_{T,\alpha}+p_{T,\beta})/2$,
  (right) the rapidity separation $\Delta \eta =
  |\eta_{\alpha}-\eta_{\beta}|$ of the charged particle pair. }
\end{figure}

\vspace{-5mm}
%%\label{}



%% References


\bibliographystyle{elsarticle-num}
\bibliography{<your-bib-database>}

\begin{thebibliography}{00}%{99}
%\bibitem{a00} D.~Kharzeev, Phys. Rev. B{\bf 633}, 260-264 (2006).
\bibitem{a05} S.~Schlichting and S.~Pratt, Phys. Rev. C {\bf 83}, 014913 (2010).
\bibitem{a02} D.~E.~Kharzeev, Phys. Rev. B{\bf 633}, 260-264 (2006).
\bibitem{a03} S.~A.~Voloshin, Phys. Rev. C{\bf 70}, 057901 (2004).
\bibitem{a04} B.~I.~Abelev {\it et al.}  [STAR Collaboration],
  %``Azimuthal Charged-Particle Correlations and Possible Local Strong Parity Violation,''
  Phys.\ Rev.\ Lett.\  {\bf 103}, 251601 (2009).
\bibitem{a09} D.~Teaney and L.~Yan, Phys. Rev. C {\bf 83}, 064904 (2011).
\bibitem{a10} S.~Pratt, S.~Schlichting and S.~Gavin, Phys. Rev. C {\bf 84}, 024909 (2011).
\bibitem{a07} B.~I.~Abelev {\it et al.}  [ALICE Collaboration],
  arXiv:1207.0900 [nucl-ex].

\bibitem{ALICE-PPR}  ALICE Collaboration, J. Phys. G:  Nucl. Part. Phys. 30 1517 (2004)
\bibitem{a06} Y.~Hori, T.~Gunji, H.~Hamagaki and S.~Schlichting arXiv:1208.0603 [nucl-th].
%\bibitem{a08} D.~E.~Kharzeev, L.~D.~McLerran and H.~J.~Warringa,
%  Nucl. Phys. A {\bf 803}, 227 (2008).
\bibitem{Ante-QM2012-proceedings}  A.~Bilandzic [ALICE Collaboration],
  these proceedings. 
\bibitem{S.Voloshin-QM2012-proceedings} S.~A.~Voloshin [ALICE Collaboration],
  these proceedings. 




\end{thebibliography}






% \end{thebibliography}

%\bigskip
%\bigskip

%\section*{References}

\end{document}